\documentclass[twocolumn,showpacs,amsmath, amssymb, prd, natbib]{revtex4-1}

\usepackage[utf8]{inputenc}
\usepackage{graphicx}
\usepackage{amssymb}
\usepackage{amsmath}
\usepackage{natbib}
\usepackage[capitalise]{cleveref}
\usepackage{color}

\newcommand{\be}{\begin{equation}}
\newcommand{\ee}{\end{equation}}
\newcommand       \bea          {\begin{eqnarray}}
\newcommand       \eea          {\end{eqnarray}}
\newcommand       \apjl         {ApJL}
\newcommand       \aap          {A\&A}
\newcommand       \aaps          {A\&A Supplements}
\newcommand       \mnras        {MNRAS}

\newcommand      \apjs {ApJ Supplements}

\def\red#1{\textcolor{red}{\textbf{}}}
\def\blue#1{#1}

\begin{document}

\title{Revisiting Stellar Orbits and the Sgr A$^*$ Quadrupole Moment}
\author{Yael Alush, Nicholas C. Stone}
\affiliation{Racah Institute of Physics, The Hebrew University, 91904, Jerusalem, Israel}
\email{yael.alush@mail.huji.ac.il}
%\date{\today}

\begin{abstract}
The ``no-hair'' theorem can, in principle, be tested at the center of the Milky Way by measuring the spin and the quadrupole moment of Sgr A$^*$ with the orbital precession of S-stars, measured over their full periods. Contrary to the original method, we show why it is possible to test the no-hair theorem using observations from only a single star, by measuring precession angles over a half-orbit. 
There are observational and theoretical reasons to expect S-stars to spin rapidly, and we have quantified the effect of stellar spin, via spin-curvature coupling (the leading-order manifestation of the Mathisson-Papapetrou-Dixon equations), on future quadrupole measurements.  \red{We find that they will typically suffer from errors of order a few percentage points, but for some orbital parameters, the error can be much higher.} \blue{We
find that spin-curvature coupling is generally a minor effect that causes errors only of order a few percentage points, but for some orbital
parameters, the error can be much higher.}
We re-examine the more general problem of astrophysical noise sources that may impede future quadrupole measurements, and find that a judicious choice of measurable precession angles can often eliminate individual noise sources.  We have derived optimal combinations of observables to eliminate the large noise source of mass precession, the novel noise of spin-curvature coupling due to stellar spin, and the more complicated noise source arising from transient quadrupole moments in the stellar potential. 
\end{abstract}

\date{\today}
\maketitle

\section{Introduction}
\label{sec:intro}
At the center of our galaxy, there is a supermassive black hole (SMBH), known as Sgr A$^*$. It is surrounded by a dense cluster of stars, the ``S-stars'' \citep{Krabbe+1995,Genzel+1997,Figer+2000,Gezari+2002,Ghez+2003}. Some of them orbit the SMBH with small semimajor axes and high eccentricities. The main sequence star S2 is the canonical example of this: it has a highly elliptical orbit with a 16 year period \citep{Gillessen+2017}. At its pericenter distance of $\approx$120AU from Sgr A$^*$, it has an orbital speed of $\approx7650$ km s$^{-1}$. The discovery of these stars has given observers the ability to test relativistic effects around a rotating black hole. For example, the Schwarzschild precession of the S2 orbit \citep{GravityCollaboration2020} and its gravitational redshift \citep{GravityCollaboration2018,Do+2019} have already been detected.

Further observations on the orbits of S-stars would allow us to test the ``no-hair'' theorem \citep{Will2008,Merritt+2010}. The no-hair theorem states that any black hole solution can be completely characterized by only three parameters: its mass $M_\bullet$, angular momentum $\mathbf{J_\bullet}=\boldsymbol{\chi_\bullet}\left(\frac{GM_\bullet^2}{c}\right)$, where $\boldsymbol{\chi_\bullet}$ is the dimensionless spin, and its electric charge. A direct consequence is that all higher multipole moments of an astrophysical (i.e. electrically neutral) black hole can be expressed as a function of only $M_\bullet$ and $\mathbf{J_\bullet}$ \citep{Geroch1970,Hansen1974}. In particular, the quadrupole moment is $Q_{2\bullet}=-\frac{1}{c}\frac{J_\bullet^2}{M_\bullet}$. To test the no-hair theorem, we need to determine five parameters: the mass of the black hole, the magnitude and two angles \footnote{In principle, testing the no-hair theorem only requires the magnitude of the SMBH spin vector, but in practice, the measurable consequences of non-zero $J_\bullet$ and $Q_{2\bullet}$ both depend on the spin vector's orientation.} of its spin, and the value of its quadrupole moment, and then verify or refute the above relationship.

For a non-rotating black hole, Schwarzschild precession is the most important relativistic effect, leading to a shift in the stars' pericenter angle. If the black hole is rotating, then more relativistic phenomena affect the S-star orbits. The Lense–Thirring (LT) effect and torques from the quadrupole moment will lead to additional apsidal precession, and are also the leading-order sources of nodal precession.
Using measurements of the S2 orbital period, observers have already constrained the mass of Sgr A$^*$ to $M_\bullet\approx4\times10^6M_\odot$ \cite{Ghez+2008,Gillessen+2009}, and by measuring the change in the orbital orientations of two stars, it is possible to determine the remaining four parameters \citep{Will2008}. Despite the recent detection of Schwarzschild precession \cite{GravityCollaboration2020}, it will be challenging
to detect higher-order effects with S2's orbit because they fall off quickly with distance from the SMBH. To detect the spin and the quadrupole moment of the SMBH, closer stars are needed, both to yield observationally detectable precession angles and also to overcome sources of noise such as gravitational perturbations from other stars \citep{Merritt+2010}. In principle, if spectroscopy can obtain radial velocity measurements, these could be combined with astrometric precession measurements to yield better constraints \citep{Waisberg+2018}.  As with astrometry, radial velocity will be most sensitive to general relativistic effects for stars closer to Sgr A$^*$.  There is thus an effort to search for stars at smaller radii and smaller orbital periods.  

However, since S-stars are so hard to find, our constraints on general relativity (GR) will often be dominated by the single most relativistic star known. Therefore, we should understand how we can test the no-hair theorem with just one star. A recent paper suggests a way to test the no-hair theorem using one star by a Markov chain Monte Carlo method using future measurements of stellar orbits \citep{Qi2021+}. \blue{The authors show that existing S2 astrometric measurements cannot constrain the spin and the quadrupole moment of the SMBH. }

Recently, a few stellar candidates with shorter periods and/or pericenters than S2 were detected \cite{Peissker+2020}. Despite the controversy over their orbital properties \citep{GravityCollaboration2021,Peissker+2021}, they may be better probes for testing relativistic effects than S2\blue{ \citep{Iorio2020}}.  Observation suggests that at least one of them, S4711, is fast-rotating with a projected rotation velocity of $V\sin{i}=239.60\pm25.21~{\rm km~s}^{-1}$. This result and similar observations \citep{Habibi+2017} indicate that many S-stars spin rapidly.  Even if the new S-star orbital solutions ultimately prove to be incorrect, ongoing upgrades to the GRAVITY instrument (``GRAVITY+'') are likely to greatly expand our sample of S stars in the near future \citep{GravityCollaboration2022}, raising the prospect of finding significantly more relativistic stellar orbits.

The spin of a test particle in a gravitational field will cause deviations from geodesic motion. Its motion can be described using the more complex Mathisson-Papapetrou-Dixon (MPD) equations \citep{Dixon1970}. Those deviations would complicate the testing of the no-hair theorem by adding a new source of noise to the orbital precession measurements.  Both this and previously considered noise sources (such as stellar perturbations) need to be carefully considered in any future tests of the no-hair theorem before discovery of an anomalous SMBH quadrupole moment can be claimed.

\red{We note that, aside from orbital measurements of the S-stars, other theoretical tests for the no-hair theorem exist \citep{Johannsen2016}. If a pulsar is located sufficiently close to Sgr A$^*$, its radio pulses could provide another means to test the no-hair theorem \citep{Wex+1999,Liu+2014}. Another approach is using gravitational-wave (GW) measurements \citep{Ryan1997,Glampedakis+2006,Saleem+2021}. Using today's Earth-based LIGO-Virgo-KAGRA detectors, it is challenging to measure an individual object's spins and quadrupole moments in a GW binary. However, in the future, using the space-based {\it LISA} detector, we will be able to detect the GWs from extreme mass ratio inspirals and calculate the multipole moments of the central BH. An alternative possibility to test the no-hair theorem is by using images of the BH ``shadows'' \citep{Johannsen+2010,Broderick+2014}. For a Schwarzschild BH, the shadow is exactly circular and centered on the BH, and for a rotating BH, the shadow is displaced but remains nearly circular (except for high spin values or large inclination). However, if the no-hair theorem is violated, the shape of the shadow can be significantly different. Fitting models to different features of accretion disks can be another way to test the no-hair theorem. This may be done on the shape of relativistically broadened iron lines or on the X-ray thermal continuum spectra \citep{Bambi+2016,Dovciak+2004,Li+2005}. In some cases, combining multiple methods will lead to a more accurate test for the no-hair theorem \citep{Psaltis+2016}, though in this paper we restrict our attention to tests using S-star orbits. } 

This paper is organized as follows. In \S\ref{sec:half-shifts} we 
analytically derive a method for testing the no-hair theorem using only one S-star. \S\ref{sec: sources of noise} describes the sources of noise due to stellar perturbations and spin-curvature coupling on the quadrupole moment measurements. \S\ref{sec: minimizing the errors} explores how those errors can be minimized, and shows that many broad categories of ``noise sources'' can be precisely removed with carefully tailored combinations of observables. In \S\ref{sec: discussion} we conclude and discuss future observations. 

\section{Shifts and Half-Shifts} \label{sec:half-shifts}
Using orbital perturbation theory, we can calculate the precessions per orbit of a star's Euler angles (i.e. orbital elements). We call these per-orbit precessions  ``full-shifts'', and to 2nd post-Newtonian order, in the extreme mass-ratio limit, they are given by \footnote{See e.g. Ref. \citep{Will2008}\blue{, and \citep{Iorio2011}} for all of these aside from the 2PN monopole term.}: 
\begin{subequations}
    \begin{align}
        \delta\varpi &= A_{\rm S}-2 A_{\rm J} \cos{\alpha}-\frac{1}{2}A_{\rm Q_2}\left(1-3\cos^2 \alpha\right)\\
        &+\left(\frac{14-e^2/2}{36\pi}\right)A_{\rm s}^2 \notag \\
        \sin{i}\delta\Omega&=\sin{\alpha}\sin{\beta}\left( A_{\rm J}-A_{\rm Q_2}\cos{\alpha} \right)\\
        \delta i&=\sin{\alpha}\cos{\beta}\left(A_{\rm J}-A_{\rm Q_2}\cos{\alpha}\right)
    \end{align}
    \label{eq:full-shifts}
\end{subequations}
where
\begin{subequations}
\begin{align}
        A_{\rm S}&=\frac{6\pi}{c^2}\frac{GM_\bullet}{\left(1-e^2\right)a}, \\
        A_{\rm J}&=\frac{4\pi\chi_\bullet}{c^3}\left[\frac{GM_\bullet}{\left(1-e^2\right)a}\right]^{3/2}, \\ 
        \begin{split}
        A_{\rm Q_2}&=\frac{3\pi }{cM_\bullet}\frac{Q_{2\bullet}}{\left(1-e^2\right)^2a^2}\\
        &=\frac{3\pi\chi_\bullet^2 }{c^4}\left[\frac{GM_\bullet}{\left(1-e^2\right)a}\right]^2.    
        \end{split}
\end{align}
\end{subequations}
Here the three angles that precess due to relativistic effects are $\varpi$, the longitude of pericenter, $\Omega$, the longitude of ascending node, and $i$, the inclination of the orbit (each angle is defined in the ``sky plane'', i.e. with a reference plane that is normal to the observer's line of sight). The quantities $a$ and $e$ are the semimajor axis and the eccentricity of the star. The polar angles of the BH's spin with respect to the stellar orbital plane are denoted by a colatitude angle, $\alpha$, and an azimuthal angle, $\beta$.  

Measuring the full-shifts of the ascending node and the inclination of two S-stars \red{in non-degenerate} orbits would allow us to calculate the four remaining parameters $\mathbf{J_\bullet}$ and $Q_{2\bullet}$ needed to test the no-hair theorem.  \blue{We note that it is important to include the 2PN contribution to apsidal precession (i.e. the last term in $\delta \varpi$) as it will generally exceed the magnitude of $A_{\rm Q_2}$.}  In this approach, two stars are needed because there is not enough information in the full-shifts of a single star.

However, there is more information contained in the relativistic orbital motion that is hidden by a full orbit average. \red{This was recently shown in Ref. \citep{Qi2021+}, which demonstrated with a Markov chain Monte Carlo method that a single stellar orbit can in principle measure these four parameters.} \blue{This was recently exploited in Ref.  \citep{Heissel+2022} to separate between the Schwarzschild and mass precession signatures within a single orbit, to constrain the mass enclosed within the S2 orbit \cite{GRAVITYCollaboration2022b}. Also,  in Ref. \citep{Qi2021+}, it was demonstrated with a Markov chain Monte Carlo method that a single stellar orbit can in principle measure the four parameters to test the no-hair theorem. }  Here we give an explicit analytic demonstration of how this additional information is contained within a single star's orbit.  Specifically, we can use the precession completed after a half-orbit (the ``half-shifts''), which in some cases are non-degenerate with the full-shifts\blue{ (i.e., we will have more independent equations to calculate the SMBH spin and quadrupole moment)}.

In order to test the no-hair theorem using the half-shifts, we need to calculate them using orbital perturbation theory over a half-orbit (specifically, from pericenter to apocenter). Using the Gauss planetary equations to integrate leading-order post-Newtonian accelerations \citep{Faye+2006,Blanchet2006}, we find:
\begin{subequations} 
    \label{eq: half-shifts}
    \begin{align}
        \delta\varpi_{\frac{1}{2}}&=\frac{\delta\varpi}{2}-\frac{1+2e^2}{3\pi e}A_{\rm Q_2}\sin^2\alpha\sin{\left(2\beta\right)},\\
        \sin{i}\delta\Omega_{\frac{1}{2}}&=\frac{\sin{i}\delta\Omega}{2}-\frac{2e}{3\pi}A_{\rm Q_2}\sin{\alpha}\cos{\alpha}\cos{\beta},\\
        \begin{split}
            \delta i_{\frac{1}{2}}&=\frac{\delta i}{2}+\sin{\alpha}\sin{\beta}\Bigl(\frac{e}{\pi}A_{\rm J}-\frac{2e}{3\pi}A_{\rm Q_2}\cos{\alpha}\Bigr).
        \end{split}
    \end{align}
\end{subequations}

\blue{Combining the half-shifts in \cref{eq: half-shifts} with the full-shifts in \cref{eq:full-shifts}, we have a set of six observables, of which five of them are independent of each other. }
\red{The half-shifts in \cref{eq: half-shifts} are mostly \footnote{Specifically, the set of six post-Newtonian half-shifts and full shifts contains five independent equations.} non-degenerate to the full-shifts.} Therefore, there is enough information encoded in the orbit of a single star to test the no-hair theorem without the need for a second star to break degeneracies.  Our choice of half-shifts is not a unique combination of measurables for breaking the degeneracies in the full-shifts of a single star, and for detailed tests one may wish to compare future observations to large libraries of geodesic or post-Newtonian orbits around central objects with arbitrary quadrupole moments.  The advantage of the approach here is its simplicity and transparency, although even using only analytic combinations of observables, one could choose differently (e.g. ``quarter-shifts'').  We limit ourselves to half-shifts in this work because (i) for eccentric orbits, the vast majority of precession happens near pericenter; (ii) further subdividing the orbit near pericenter will increase the statistical errors on any real observation; (iii) our definition of half-shifts (i.e. integrating the true anomaly from 0 to $\pi$) produces the simplest analytic form that still breaks full-shift degeneracies.  As we will see, an additional benefit of the analytic approach taken here is that we can design combinations of observables that, by construction, cancel out astrophysically relevant sources of noise.

\section{Sources of Noise} \label{sec: sources of noise}

\subsection{Stellar Perturbations}
Previous studies \citep{Merritt+2010} showed that the presence of other stars in the cluster around the SMBH can induce orbital precession at the same order of magnitude as relativistic effects.  For future observations aimed at testing the no-hair theorem (or other aspects of GR), these stellar perturbations are a noise source whose relative importance increases with distance from the SMBH.

To a first-order approximation, the stellar distribution can be approximated as a smooth spherical cluster with a mass density $\rho\propto r^{-\Gamma}$, where $r$ is the distance from the SMBH. The spherical component of the gravitational field causes apsidal precession, such that the shift and the half-shift of the pericenter are\blue{ \cite{Merritt+2010}}:
\begin{subequations}
\label{eq:massprec}
    \begin{align}
        \delta\varpi_{\rm mass}& = 2\pi\frac{M_\star\left(a\right)}{M_\bullet}\sqrt{1-e^2}F\left(\Gamma,e\right) \label{eq:massprecfull}\\
        \delta\varpi_{\rm \frac{1}{2},mass}&=\frac{\delta\varpi_{\rm mass}}{2} \label{eq:massprechalf},
    \end{align}
\end{subequations} 
where $M_{\star}(a)$ is the mass enclosed within radius $r=a$, and $F\sim 1$ is a weak function of $e$ and $\Gamma$ ($F=1$ exactly for $\Gamma=1$) \citep{Merritt+2010}. This ``mass-precession'' effect can mimic relativistic precession and cause error in the spin and the quadrupole moment measurements. \blue{We note that the mass precession dominates in the apocenter while the GR effects dominate in the pericenter \cite{Heissel+2022}.}

Non-spherically symmetric perturbations, such as vector resonant relaxation (VRR), can also create a source of error by changing the orientation of the orbital planes. Unlike the deterministic effect of mass precession, VRR is usually modeled as a stochastic perturbation to the longitude of the ascending node\blue{ \cite{Merritt+2010}}: 
\begin{equation}
    \delta\Omega_{\rm VRR}\sim q\sqrt{N}
\end{equation}
where $q=m_{\star}/M_{\bullet}$ and $N$ is the number of stars inside the orbit of the test star.  However, this picture of VRR is only relevant over long (secular) timescales, considering many orbit-averaged interactions between different stars.  Fundamentally, VRR is driven by the stochastically varying net multipole moments of the total stellar potential, each of which emerge as a result of Poissonian discreteness in the stellar population.  Over one to a few dynamical times, however, these multipole moments do not have time to evolve and can be regarded as ``frozen in''.  For the purposes of measuring S star orbital precession, therefore, we will estimate the statistically typical values of these multipole moments using the formalism of Ref. \citep{KocsisTremaine2015}.

The stellar potential's quadrupole $A_{\rm Q_2^\star}$, the lowest order aspherical contribution in the multipole expansion, dominates over higher multipole moments, which combined contribute about $\approx 10\%$ as much precession as $A_{\rm Q_2^\star}$ \citep{KocsisTremaine2015}. Therefore, in this paper we will only consider the leading-order multipole moment, $A_{\rm Q_2^\star}$. The precessions due to the stellar quadrupole moment are:
\begin{subequations} 
    \label{eq:stellarquad}
    \begin{align}
       \delta\varpi_{\rm Q_2^\star} &=-\frac{1}{2}A_{\rm Q_2^\star}\left(1-3\cos^2 \alpha^\star\right)\\
        \sin{i}\delta\Omega_{\rm Q_2^\star}&=-A_{\rm Q_2^\star}\cos{\alpha^\star}\sin{\alpha^\star}\sin{\beta^\star}\\
        \delta i_{\rm Q_2^\star}&=-A_{\rm Q_2^\star}\cos{\alpha^\star}\sin{\alpha^\star}\cos{\beta^\star}\\
        \delta\varpi_{\rm \frac{1}{2},Q_2^\star}&=\frac{\delta\varpi_{\rm Q_2^\star}}{2}-\frac{1+2e^2}{3\pi e}A_{\rm Q_2^\star}\sin^2\alpha^\star\sin{\left(2\beta^\star\right)},\\
        \sin{i}\delta\Omega_{\rm\frac{1}{2},Q_2^\star}&=\frac{\sin{i}\delta\Omega_{\rm Q_2^\star}}{2}-\frac{2e}{3\pi}A_{\rm Q_2^\star}\sin{\alpha^\star}\cos{\alpha^\star}\cos{\beta^\star},\\
        \delta i_{\rm \frac{1}{2},Q_2^\star}&=\frac{\delta i_{\rm Q_2^\star}}{2} -\frac{2e}{3\pi}A_{\rm Q_2^\star}\cos{\alpha^\star}\sin{\alpha^\star}\sin{\beta^\star}
    \end{align}
\end{subequations}
where $\alpha^\star$ and $\beta^\star$ are the polar angles of the stellar quadrupole with respect to the stellar orbital plane.  The deterministic nature of precession due to the stellar quadrupole moment contrasts strongly with the usual stochasticity of VRR, and we emphasize that this difference arises merely because of the very short timescales relevant for no-hair tests.  This determinism is actually quite advantageous; in \S \ref{sec: minimizing the errors}, we will show that it permits us to precisely eliminate this noise source. 

\subsection{Spin-Curvature Coupling} 
As mentioned in \S \ref{sec:intro}, there is substantial 
direct evidence \citep{Habibi+2017} that S-stars spin rapidly. Moreover, there are empirical and theoretical reasons to expect S-stars to rotate rapidly. First, the S-stars are B stars. In the field, many B-type stars spin with an equatorial rotation velocity of about $250 ~{\rm km~s}^{-1}\sim30\%$ of the centrifugal breakup limit \citep{Dufton+2013}. Second, the S-stars are located in a dense cluster around a SMBH. The dense environment leads to close, repeated hyperbolic tidal encounters between the stars, which spins them up \citep{Alexander+2001}. A similar effect may lead to tidal spin-up of stars during close passages near the SMBH \citep{Goicovic+2019,Ryu+2020}.  If in the future, a rapidly spinning S-star is used for no-hair tests, it will be important to understand the additional precessions due to the MPD equations (of which spin-curvature coupling is the leading-order effect) as an additional noise source.

To quantify the importance of spin-curvature coupling, we need to calculate the precessions of the orbital elements due to the MPD effect. We denote the mass of the star by $m_\star\ll M_\bullet$, its dimensionless spin magnitude by $s_\star=S_\star\left(Gm_\star^2/c\right)^{-1}$ (here $S_\star$ is its dimensional angular momentum), and its direction with two polar angles, a colatitude angle, $\gamma$, and an azimuthal angle, $\delta$, both relative to the star's orbital plane.

Before calculating the MPD shifts, we need to check how fast the star's spin itself precesses during its motion. To estimate how much the spin direction changes during one orbital period, we integrated the leading-order post-Newtonian (PN) approximation for the rate of stellar spin precession \citep{Faye+2006, Racine2008}, i.e. the geodetic precession.
The spin direction changes over time, but only very slowly. The spin of the star will return to its initial direction after $\sim a\left(1-e^2\right)\left(GM_\bullet/c^2\right)^{-1}=(a/r_{\rm g}) \left(1-e^2\right)$ orbital periods (here $r_{\rm g}=GM_\bullet/c^2$ is the gravitational radius). This conclusion is also shown in previous research \citep{Fang+2021} in a more detailed examination, using higher orders of PN spin precession. Therefore, we approximate the spin direction as fixed in the following calculations, greatly simplifying the orbit-averaging procedure.

Orbital perturbations due to the rotation of a star appear at lowest-order in the $1.5{\rm PN}$ term of the PN approximation \citep{Faye+2006}\blue{, and after orbit-averaging, the shifts are given by:}\red{. The shifts and the half-shifts due to the MPD effect are:} 
\begin{subequations}
\label{eq:MPD}
    \begin{align}
        \delta\varpi_{\rm MPD}&=-6A_{\rm MPD}\cos{\gamma},\\
        \sin{i}\delta\Omega_{\rm MPD}&=3A_{\rm MPD}\sin{\gamma}\sin{\delta},\\
        \delta i_{\rm MPD}&=3A_{\rm MPD}\sin{\gamma}\cos{\delta},\\
        \delta\varpi_{\rm \frac{1}{2},MPD}&=\frac{\delta\varpi_{\rm MPD}}{2} \label{eq:pericenter half shift MPD}, \\
        \begin{split}
        \sin{i}\delta\Omega_{\rm\frac{1}{2},MPD}&=\frac{\sin{i}\delta\Omega_{\rm MPD}}{2}\\&+\frac{2e}{\pi}A_{\rm MPD}\sin{\gamma}\cos{\delta},  
        \end{split}\\
        \delta i_{\rm \frac{1}{2},MPD}&=\frac{\delta i_{\rm MPD}}{2}+\frac{2e}{\pi}A_{\rm MPD}\sin{\gamma}\sin{\delta}
    \end{align}
\end{subequations}
where
\begin{equation}
    A_{\rm MPD}=\frac{\pi s_\star}{c^3}\frac{m_\star}{M_\bullet}\left[ \frac{GM_\bullet}{\left(1-e^2\right)a}\right]^{3/2}.
\end{equation}

\begin{figure}
\includegraphics[width=80mm]{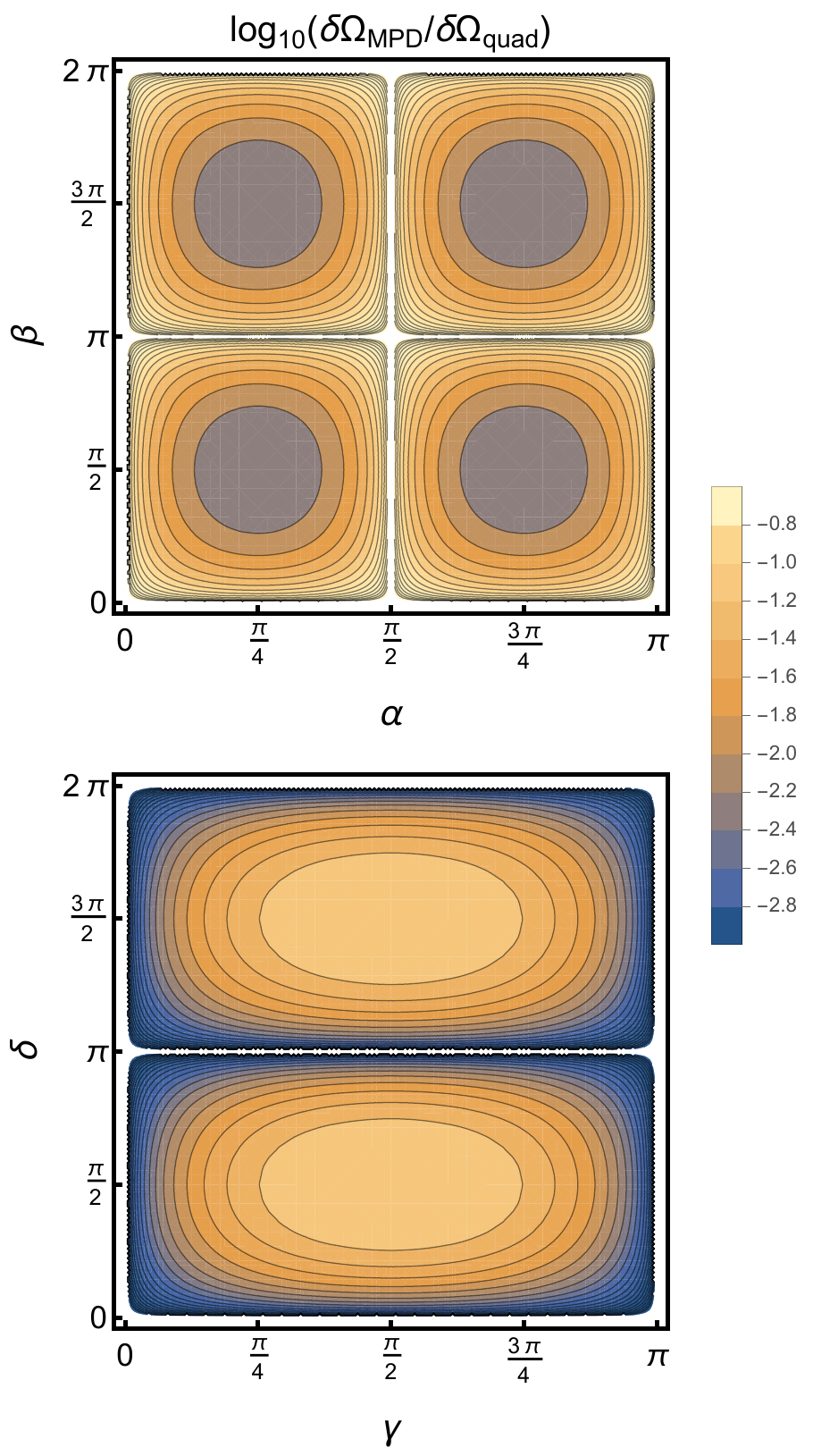}
\caption{The ratios between the MPD and the quadrupole precessions as a function of the spins' directions. We show the results with representative values for the angles. {\it Top}: The spin orientation of the star is taken to be $\gamma=\pi/3$, $\delta=\pi/5$. {\it Bottom}: The spin orientation of the SMBH is taken to be $\alpha=\pi/8$, $\beta=\pi/6$. The SMBH spin and mass are $M_\bullet=4\times10^6M_\odot$ and $\chi_\bullet=0.5$. The mass and the radius of the star are $m=10M_\odot$ and $r=\sqrt{10}R_\odot$. The star spins with $30\%$ of its breakup frequency, and orbits the SMBH with a semimajor axis of $a=3~{\rm mpc}$, and eccentricity of $e=0.8$. The color bar presents the ratio $\delta \Omega_{\rm MPD} / \delta\Omega_{\rm quad}$ in a base-10 logarithmic scale.}
\label{fig:ratios vs. spins directions}
\end{figure}

We are particularly interested in the influence of the MPD effect on the quadrupole measurements. Therefore, we show in \cref{fig:ratios vs. spins directions} the ratios between the MPD precessions and the quadrupole precessions, as functions of different angles (the stellar orbit is chosen to resemble that of S2). In both cases, the spin directions of the SMBH and the star can affect the ratios by almost two orders of magnitude. 

\blue{
\subsection{Tidal Force}
At very small pericenters, tidal interactions between the SMBH and the star can cause a level of precession that overwhelms the higher order GR shifts we are interested in. The equilibrium tide describes quasi-static changes to the shape of an extended object due to a slowly varying tidal field acting on it. The force of the equilibrium tide is in the radial direction, and it thus only causes an apsidal precession \cite{Fabrycky+2007}:
\begin{equation}
    \delta\varpi_{\rm tide}=\frac{15\pi M_\bullet}{4a^5m_\star}\frac{8+12e^2+e^4}{(1-e^2)^5}kr_\star^5
\end{equation}
where $m_\star$ and $r_\star$ are the mass and the radius of the star, respectively, and $k$ is the star's tidal Love number.} 

\blue{The dynamical tide describes the more general situation where the tidal deformations are not necessarily quasi-static. It is usually relevant only for pericenters near the tidal radius of the star. The local apsidal precession rate due to the dynamical tide is \citep{Kumar1998}:
\begin{equation}
    \frac{\dot{\varpi}_{\rm dyn\, tide}}{2\pi}\sim\frac{\dot{L}}{L}\approx\frac{\dot{E}\Omega_p}{L}
\end{equation}
where $L$ is the star's orbital angular momentum, $E$ is the orbital energy, and $\Omega_p$ is the orbital angular frequency of the star at pericenter. For a highly eccentric orbit, most of precession happens near pericenter, so the shift is:
\begin{equation}
    \delta\varpi\sim\frac{\Delta E}{L}\Omega_p
\end{equation}
where $\Delta E$ is the change in the orbital energy per orbit. The change in $E$ can be expressed by \citep{Lee+1986}:
\begin{equation}
    \Delta E=\left(\frac{Gm_\star}{r_\star}\right)^2\left(\frac{M_\bullet}{m_\star}\right)^2\sum_{l=2,3,...}\left(\frac{r_\star}{r_p}\right)^{2l+2}T_l \left(\frac{r_p}{r_\star}\right)
\end{equation}
where $l$ is the spherical harmonic index, $r_p$ is the pericenter distance, and the function $T_l$ falls off exponentially with distance from the SMBH. Therefore, the dynamical tides are negligible in our case.
}

\section{Minimizing the Errors}\label{sec: minimizing the errors}
\begin{figure*}
\centering
\includegraphics[width=160mm]{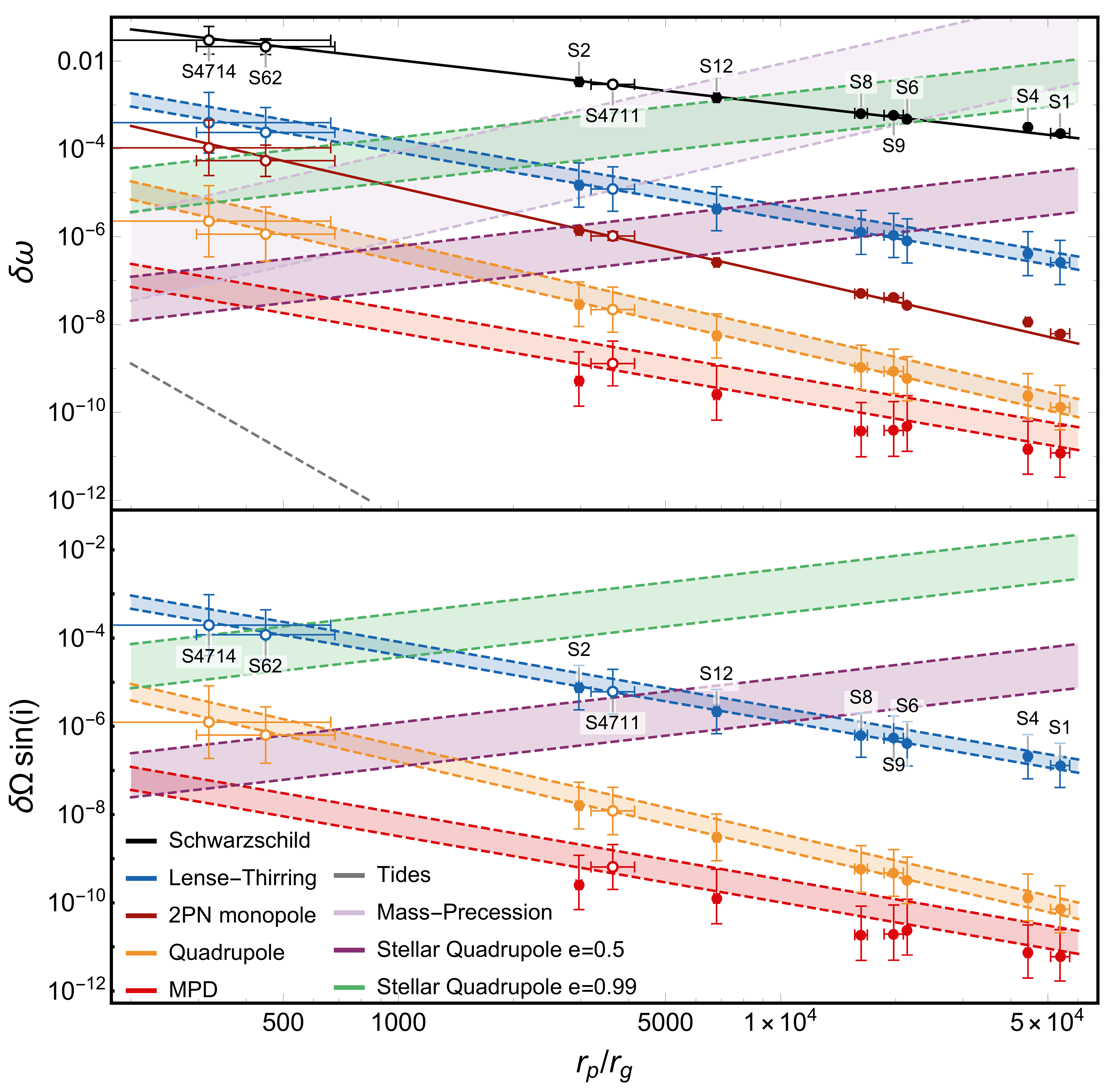}
\caption{Full-shifts plotted against the dimensionless pericenter distance, with different effects color-coded as per the label in the bottom panel. {\it Top}: The full-shift of the argument of the pericenter. {\it Bottom}: The full-shift of the longitude of the ascending node. The mass and the spin of the SMBH are $M_\bullet=4\times10^6M_\odot$, $\chi_\bullet=0.5$. The S-star mass is always taken to be $m=10M_\odot$, and its eccentricity is $e=0.8$ except when otherwise noted. The boundaries of the shaded areas of the LT and the quadrupole terms are the maximum shifts and the isotropic averages, given uncertainties in $\alpha$ and $\beta$. The mass precession and the stellar quadrupole are presented for a number density exponent $\Gamma=1$, and their boundaries are for distributed masses of $1M_\odot$ and $100M_\odot$ enclosed within $1~{\rm mpc}$. The stellar quadrupole direction is taken to be $\alpha^\star=\pi/4$, $\beta^\star=\pi/4$. The boundaries of the MPD effect are for the breakup spin and $30\%$ of it. The data points are the shifts for different S-stars with different observed spin magnitudes. Data points are shown for various S-stars with small pericenters; the center of each data point is shown at the isotropic mean of the relevant unknown angular variables, and the error bars are a sum of the isotropic variances and the observational uncertainties of the stellar orbits \citep{Gillessen+2017}. The spin magnitude for each data point is the observed spin \citep{Habibi+2017}. The tidal precession \citep{Fabrycky+2007} is presented for a stellar Love number $k=0.014$, approximating the B-type star as an n=3 polytrope \citep{Yip+2017}.}
\label{fig:shifts vs. pericenter}
\end{figure*}

Now that we have the mathematical description for the mass precession, the stellar quadrupole, and the MPD shifts, we can compare these effects and see how the existence of other stars, and the spin of the test star itself affect the measurements. As we will see, the simple functional form of most sources of noise will allow us to minimize these errors with careful choice of observables.  In \cref{fig:shifts vs. pericenter}, we present the apsidal and nodal precessions as a function of the dimensionless pericenter distance due to different effects.  
 
In our results, the mass of the SMBH is $M_\bullet=4\times10^6 M_\odot$, the mass of Sgr A$^*$, and the spin magnitude is $\chi_\bullet=0.5$. Most of the detected S-stars are massive, so we choose to present the shifts for a star with mass $m_\star=10M_\odot$ and a radius $r_\star=10^{0.5}R_\odot$ \citep{Kippenhahn+2012}. However, we assume that selection effects have so far prevented the detection of a larger population of fainter stars. Therefore, in this figure, the mean stellar mass (of the population of background stars) is $1M_\odot$. We use a tabulated moment of inertia $I$ \citep{Claret+1989} for the stellar breakup frequency because real stars are centrally concentrated. High-mass stars that are on the main sequence have $I\sim 0.09m_\star r_\star^2$. Moreover, we take empirically measured spin magnitudes \citep{Habibi+2017}. However, we do not know the true orientation of the stellar spins. Therefore, to get a typical value from observations of the projected spin, we need to take an ``isotropic average''- we assume that the probability for each angle between the observed spin magnitude to the breakup spin is the same, and then we take an average angle. We will also use the concept of the isotropic average when showing purely theoretical predictions that depend on unknown angle variables ($\alpha$, $\beta$, $\gamma$, $\delta$).

\blue{Our estimates of both mass precession and precession due to the stellar quadrupole moment require making assumptions on the unresolved stellar mass distribution at small radii. For concreteness, we take $\Gamma=1$ and consider a range of masses enclosed within $1~{\rm mpc}$, from $1 M_\odot$ (a very low value) to $100 M_\odot$, a value close to the upper limit inferred by Ref. \citep{GRAVITYCollaboration2022b}.}

We can see that all shifts due to GR effects fall off as a negative power of the dimensionless pericenter distance $r_{\rm p}/r_{\rm g}$. Conversely, the stellar perturbations weaken at smaller radii. Therefore, more relativistic stellar pericenters are doubly desirable, as they both increase the relativistic shifts and decrease a major noise source. Furthermore, we note that astrometric measurements will more easily detect an angular shift (e.g. $\delta \varpi$ or $\delta \Omega$) of fixed magnitude if the star's semimajor axis $a$ is smaller \citep{Waisberg+2018}. This is because the shorter orbital period of such a star ($\propto a^{3/2}$) will accumulate a greater total shift in a fixed observing window, outweighing the competing effect of smaller angular size of the orbit ($\propto a$). \red{While at very small pericenters, tidal interactions between the SMBH and the star can cause a level of precession that overwhelms the higher-order GR shifts we care about, we show in \cref{fig:shifts vs. pericenter} that tides are highly subdominant in the radii of current interest.} \blue{When a star is getting closer to the SMBH, tidal forces are more important. However, we show in \cref{fig:shifts vs. pericenter} that tides are highly subdominant in current-interest radii. Therefore we will not consider them in our following estimations.}

In \cref{fig:shifts vs. pericenter} we see that a star with orbital properties similar to S2 will see shifts from $Q_{2\bullet}$ swamped by stellar perturbations, possibly by multiple orders of magnitude (in agreement with past work \citep{Merritt+2010, Qi2021+}).  Stars such as S4714 and S62 \citep{Peissker+2020} may have quadrupole-order shifts that dominate precession due to the background stellar potential, although as noted earlier, the orbital solutions of these stars remain contested at present \citep{GravityCollaboration2021}.  
We also notice that MPD effects due to the star's spin are almost always subdominant to precession from the SMBH quadrupole moment, although spin-curvature coupling may set a noise floor of $\sim 1-10\%$ in future no-hair tests.  We note that it is hard to determine the full spin vector of a distant star because it depends on the unknown inclination of the star's pole to the line of sight. In our calculations, we can only set lower and upper limits to the spin magnitude by measuring the projected magnitude and calculating the breakup spin, respectively.

We will now estimate more precisely how much mass precession, a quadrupole moment in the stellar potential, and the MPD effect can influence the SMBH quadrupole measurement if these sources of noise are ignored during parameter estimation. To do this systematically, we used a Monte Carlo method. We chose $10^4$ random directions for the spin of the SMBH and another $10^4$ random directions for the spin of the star, and the stellar quadrupole. For each iteration, we calculated the SMBH quadrupole moment, assuming perfect measurement of relevant shifts and half-shifts, but neglecting the effects of stellar perturbations and stellar spin when converting these observables into $Q_{2\bullet}$. We calculated $Q_{2\bullet}$ a second time adding in the influence of other stars and the MPD precession terms, and then found the relative error between the measurements. Our results are computed for test stars with the properties of S2\blue{ and the closest known star to the SMBH, S4714}: \red{its mass, radius, semimajor axis, and eccentricity.} \blue{their masses, semimajor axes, and eccentricities. For S2, we use the observed radius, but S4714's radius is unknown, so we use  $r_\star=(m_\star/M_\odot)^{0.5}R_\odot$ \citep{Kippenhahn+2012}. The spin magnitude for S2 is $23\%$ of breakup, which is the average deprojection of S2's observed value. The radial velocity of S4714 has not been measured; therefore, we present our results with a spin magnitude of $20\%$ of breakup.} The distributed mass enclosed within $1~{\rm mpc}$ is \red{$10M_\odot$} \blue{$35M_\odot$, the estimated upper limit when taking $\Gamma=1$ and extrapolating inwards from constraints on distributed mass inside the S2 apocenter \citep{GRAVITYCollaboration2022b}.}\red{, and the spin magnitude, which is $23\%$ of breakup, is the average deprojection of S2's observed value.} We note that the errors can increase or decrease for different orbital and stellar elements.  We consider three scenarios for the astrophysical noise background: (i) only MPD precession, (ii) MPD plus mass precession, (iii) MPD, mass precession, and a stellar quadrupole moment.

In order to convert mock observations into estimates of $Q_{2\bullet}$, we need to make a concrete choice of observables, which for us are different combinations of full- and half-shifts (for one or more stars).  However, the simple functional form of different astrophysical noise sources (Eqs. \ref{eq:massprec}, \ref{eq:stellarquad}, and \ref{eq:MPD}) suggests that by careful algebraic re-arrangment of these observables, we can tailor combinations of full- and half-shifts that by design will completely eliminate individual astrophysical noise sources.  

A simple example of this is removing the influence of mass precession.  Since the monopole component of the stellar potential only creates apsidal (not nodal) precession, we can eliminate this noise source altogether by using the full-shifts $\sin i \delta\Omega$ and $\delta i$ for two different stellar orbits.  If we are limited to a single star, however, we seem to have a problem.  While we need four observables to isolate $Q_{2\bullet}$, we cannot use a simple combination such as $\{\sin i \delta\Omega, \delta i, \sin i \delta\Omega_{\frac{1}{2}}, \delta i_{\frac{1}{2}} \}$, because these observables are mutually degenerate and only offer three independent measurements.  However, we can exploit the simple symmetry of the apsidal half-shift due to the mass precession ($\delta \varpi_{\rm \frac{1}{2}, mass} = \delta \varpi_{\rm mass}/2$, but $\delta \varpi_{\frac{1}{2}} \ne \delta \varpi / 2$) to construct a new observable that removes mass precession while leaving a contribution from $Q_{2\bullet}$: $\delta\varpi_{\rm sub}\equiv\delta\varpi-2\delta\varpi_{\frac{1}{2}}$ (see a nice illustration in the figures in \citep{Heissel+2022}).  We can now eliminate mass precession entirely using only shifts and half-shifts from a single star, e.g. using the combination $\{\sin i \delta\Omega, \delta i, \sin i \delta\Omega_{\frac{1}{2}}, \delta \varpi_{\rm sub} \}$.

A similar approach can be taken to eliminate the combined influence of MPD effects and mass precession.  We note that 
\begin{subequations}
\begin{align}
    \sin i \delta \Omega_{\rm MPD} - 2\sin i \delta \Omega_{\rm\frac{1}{2}, MPD} + \frac{4e}{3\pi} \delta i_{\rm MPD} =& 0 \\
    \delta i_{\rm MPD} - 2 \delta i_{\frac{1}{2},\rm MPD} + \frac{4e}{3\pi} \sin i \delta \Omega_{\rm MPD} =& 0 \\
    \delta \varpi_{\rm MPD} - 2 \delta\varpi_{\rm\frac{1}{2}, MPD} =& 0.
\end{align}
\end{subequations}
We further note that the observables on the left-hand side of the above combinations are independent of mass precession, and their post-Newtonian values are 
\begin{subequations}
\label{eq:noMPDerror}
\begin{align}
    &\sin i \delta \Omega - 2\sin i \delta \Omega_{\frac{1}{2}} + \frac{4e}{3\pi} \delta i = \frac{4e}{3\pi} A_{\rm J} \sin\alpha \cos\beta \\
    &\delta i - 2 \delta i_{\frac{1}{2}} + \frac{4e}{3\pi} \sin i \delta \Omega = -\frac{2e}{3\pi} A_{\rm J} \sin\alpha \sin\beta \\
    &\delta \varpi - 2 \delta\varpi_{\frac{1}{2}} = \frac{2(1+2e^2)}{3\pi e} A_{\rm Q_2}\sin^2\alpha \sin(2\beta).
\end{align}
\end{subequations}

In the same spirit, we can eliminate the combined effect of mass precession and precession due to the mean-field stellar quadrupole moment, $A_{\rm Q_2^\star}$.  The algebra here is somewhat more involved so we present the result and its derivation in the Appendix.

\begin{figure*}
\centering
\includegraphics[width=160mm]{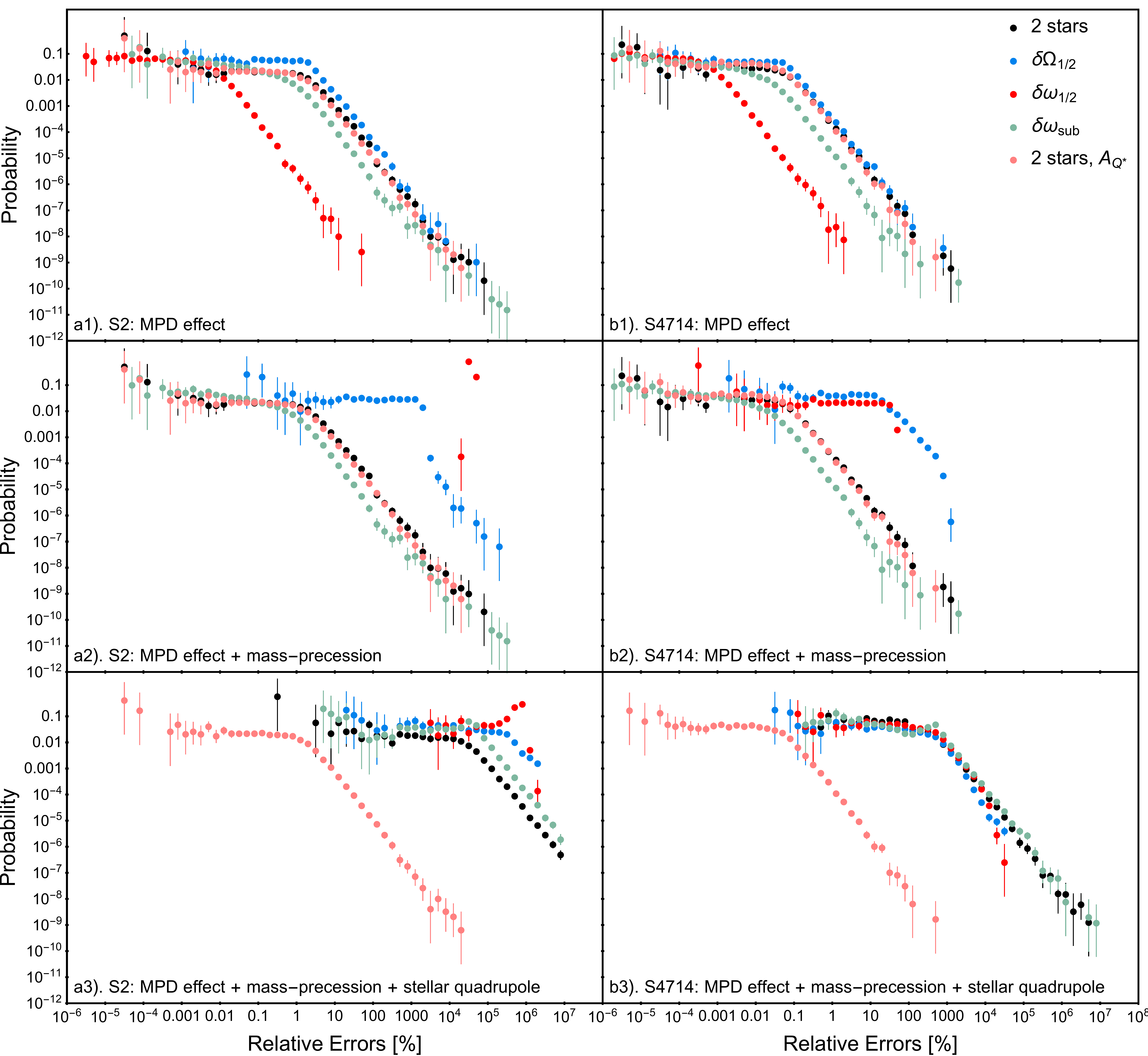}
\caption{Histograms showing relative errors in the quadrupole measurements due to {\it top}: MPD effects, {\it middle}: MPD effects and mass precession, and {\it bottom}: MPD effects, mass precession and a stellar quadrupole moment\blue{, for {\it left}: S2-like orbit, and {\it right}: S4714-like orbit (see main text)}. The results are presented for five different combinations of observables. {\it Black}: The original approach \citep{Will2008} using the full-shifts of two stars. {\it Blue}: Using three full shifts and the nodal half-shift ($\delta\Omega_{\frac{1}{2}}$) of a single star. {\it Red}: Using three full shifts and the half-shift of the pericenter ($\delta\varpi_{\frac{1}{2}}$) for a single star. {\it Green}: Using two full shifts ($\delta\Omega, \delta I$), the half-shift of the nodal angle ($\delta\Omega_{\frac{1}{2}}$), and the subtraction $\delta\varpi_{\rm sub}\equiv\delta\varpi-2\delta\varpi_{\frac{1}{2}}$ for a single star. {\it Pink}: full shifts and half-shifts of two stars, with the removal of the stellar quadrupole noise. \red{Stellar orbits and properties follow that of \red{S2} (see main text), and the} \blue{The} mass enclosed in $1~{\rm mpc}$ is \red{$10 M_\odot$}\blue{$35 M_\odot$}. The error bars were calculated using approximate analytic expressions based on Poisson statistics at the $95\%$ confidence level \citep{Gehrels86}.  
}
\label{fig:relative errors  histograms}
\end{figure*}

We now estimate the percentage errors in the calculation of $Q_{\rm 2\bullet}$ estimation for three aforementioned combinations of astrophysical noise, and five different combinations of observables:
\begin{enumerate}
    \item \label{itm:Will's approach} The original method \citep{Will2008} using the full-shifts of two stars ($\delta\Omega$ and $\delta i$). The two stars are assumed to have S2's semimajor axis and eccentricity, but with random angular orbital elements.
    \item \label{itm:nodal half shift} Using three full shifts ($\delta \varpi$, $\delta \Omega$ and $\delta i$) and the nodal half-shift ($\delta\Omega_{\frac{1}{2}}$) for a single star. 
    \item \label{itm:pericenter half shift} Using three full shifts ($\delta \varpi$, $\delta \Omega$ and $\delta i$) and the half-shift of the pericenter ($\delta\varpi_{\frac{1}{2}}$) for a single star.
    \item \label{itm:omega sub} Using two full shifts ($\delta \Omega$ and $\delta i$), the nodal half-shift ($\delta\Omega_{\frac{1}{2}}$), and the subtraction $\delta\varpi_{\rm sub}\equiv\delta\varpi-2\delta\varpi_{\frac{1}{2}}$, for a single star.
    \item \label{itm:two stars Aqs} Using shifts and half-shifts of two stars, in such a way as to remove the stellar quadrupole and mass precession noise (see Appendix).
\end{enumerate} 

Our results are presented in \cref{fig:relative errors  histograms}, which displays histograms of the relative errors in the estimated BH quadrupole moment for the five different approaches. In the top panel (MPD precession alone), the median errors for the first, second and fifth methods \blue{for S2-like orbit} are $2.3\%$, $2.2\%$, and $1.9\%$, respectively\blue{, while for S4714 they are $0.07\%$, $0.07\%$, and $0.06\%$}, though they can reach a higher value in some parts of parameter space. This example shows how the MPD effect can set a noise floor, albeit a low one, for $Q_{2\bullet}$ measurements. The fourth approach gives a moderately better result (median error of $0.75\%$ \blue{for S2 and $0.02\%$ for S4714}). However, the errors are about two orders of magnitude smaller for the third combination of shifts and half-shifts (median error of $0.007\%$ \blue{for S2 and $0.0007\%$ for S4714}). This means that we have found a surprisingly good combination of half-shifts that reduces the spin-induced error of no-hair tests by roughly two orders of magnitude compared to generic shift/half-shift combinations.  Had we used the combination of observables in. \cref{eq:noMPDerror}, there would have been exactly no error on this measurement of $Q_{2\bullet}$ (note that because \cref{eq:noMPDerror} consist of only three equations, they cannot be applied to a single stellar orbit; at least two stars are required).

Next, we added the mass precession to the MPD effect to examine the influence of other stars. We see that the error of the original approach does not change because there is no $\delta\varpi$ dependence. Conversely, we see a tremendous addition to the error in the second and third approaches due to mass precession; median errors here have risen to \red{$300\%$ and $3300\%$}  \blue{$1000\%$ and $38000\%$ for S2, and $21\%$ and $20\%$ to S4714}, respectively.  But we also see that the error in the fourth and the fifth approaches do not change, even though these both contain a $\delta\varpi$ dependence. 

To explain this contrast, the BH quadrupole moment can be solved for exactly in the third approach as: 
\begin{equation}
\label{eq:thirdapproach}
\begin{split}
A_{\rm Q_2}=&\frac{1}{2g}\Bigl(4f^2+g^2+2gA_{\rm S}-2g\delta\varpi\\
&-2f\sqrt{4f^2-2g\left(g+2\delta\varpi-2A_{\rm S}\right)}\Bigr),
\end{split}
\end{equation}
and in the fourth approach as:
\begin{equation}
A_{\rm Q_2}=\frac{1}{g}\left(h-f\right)^2+g, \label{eq:fourthapproach}
\end{equation}
where for both
\begin{subequations}
\begin{align}
        f&\equiv\sqrt{\left(\sin{i}\delta\Omega\right)^2+\left(\delta i\right)^2}, \\
        g&\equiv\frac{3\pi e}{4\left(1+2e^2\right)}\frac{f^2}{\left(\sin{i}\delta\Omega\right)\delta i}\delta\varpi_{\rm sub},\\
        h&\equiv f\left[\frac{3\pi}{2e}\frac{1}{\delta i}\left(\frac{\sin{i}\delta\Omega}{2}-\sin{i}\delta\Omega_{\frac{1}{2}}\right)+1\right].
\end{align}
\end{subequations}
In Eq. \ref{eq:thirdapproach}, $\delta\varpi$ explicitly enters the formula for $A_{\rm Q_2}$, which therefore acquires a strong dependence on mass precession.  Conversely, in Eq. \ref{eq:fourthapproach}, apsidal precession rates only appear via $\delta \varpi_{\rm sub}$, and as described earlier, this particular combination of $\delta \varpi$ and $\delta \varpi_{\frac{1}{2}}$ completely removes the effect of mass precession due to symmetry.

Finally, in the bottom panel of Fig. \ref{fig:relative errors  histograms}, we consider MPD precessions, mass precession, and precession from the stellar quadrupole moment all simultaneously.  We see that the first four combinations of observables we consider all massively mis-estimate $A_{\rm Q_2}$, typically by $2-3$ orders of magnitude: median errors for approaches 1, 2, 3, and 4 are, respectively, \red{$7300\%$, $18000\%$, $8400\%$, and $90000\%$} \blue{$30000\%$, $300000\%$, $700000\%$, and $50000\%$ for S2, and $400\%$, $450\%$, $500\%$, and $600\%$ for S4714}. For approach number 3 the error is so large \blue{when using the S2's parameters,} that we even get a large fraction of complex values (more than $50\%$). The reason for this can already be seen in Fig. \ref{fig:shifts vs. pericenter}: precession due to the stellar quadrupole moment dominates precession due to the BH quadrupole moment by $\approx 2$ orders of magnitude for S2-like stars.  Notably, however, our fifth approach (see Appendix) still produces a precision measurement of $A_{\rm Q_2}$ despite the inclusion of this additional noise source, retaining a median precentage error of $1.9\%$ \blue{for S2, and $0.06\%$ for S4714}.  This reflects the power of tailor-made observable combinations that are capable of removing deterministic sources of noise; this approach was constructed to completely remove the effects of mass precession and the stellar quadrupole moment, and all remaining error is due to the (comparatively weak) MPD effects.

\section{Discussion} \label{sec: discussion}
General relativity predicts the no-hair theorem, which states that astrophysical BHs are fully characterized by their masses and spins, and are described by the Kerr metric. The discovery of the S-stars at the center of our Galaxy allows us the opportunity to probe the curved spacetime of a rotating BH and verify whether Sgr A$^*$ is a BH of the type predicted by classical GR. Previous analytic work showed that it is possible to test the no-hair theorem by combining the orbital precession measurements from two S-stars, taken over their full orbital periods \citep{Will2008}, and recent work showed numerically that this type of test could be extended even to the orbit of a single star \citep{Qi2021+}. 

However, with current instruments, we cannot detect such high-order effects with the stars we already observe. Even using two new S-stars with more relativistic orbits that may have been recently discovered \citep{Peissker+2020}, we would have to monitor their orbits for at least $\approx 20$ years in order to detect a spin of $\chi_\bullet=0.9$ \citep{Waisberg+2018}. If those stars are real, they offer a good chance at measuring the Sgr A$^*$ spin using existing instrumentation. The quadrupole moment is probably beyond the reach of current infrared optics technology and will need to wait for improvements in resolution and/or sensitivity \citep{GravityCollaboration2022}. 

Therefore, we hope that new faint stars with more relativistic orbits will be observed. Previous work \citep{Waisberg+2018} estimated the expected number of stars, for which the GRAVITY instrument would be able to detect the spin of the SMBH. For a dimensionless spin of $\chi_\bullet=0.9$ the expected number is $0.035$ and $0.12$ for a 4-year and 10-year observing campaign, respectively. The next generation of this instrument, the GRAVITY+ project, would be improved in sensitivity to $K=22 ~{\rm mag}$. With this improvement, we will be able to detect fainter stars in more relativistic orbits, and would increase the expected number of stars by a factor of 4.

In this paper, we explain physically how it is possible to test the no-hair theorem using observations from only one star with a simple analytical method. A single star orbit is sufficient if one considers precessions seen over partial orbits (in this paper, we have examined the simple case of half-orbits). There is extra information thrown away in the full orbital average that is accessible considering fractions of orbits. We also showed that the high expected spins of the S-stars may perturb precession measurements due to spin-curvature coupling (the leading-order MPD effect). We quantified the effect of the stellar spin on the quadrupole measurements and found that for most cases, the relative errors are of order a few percentage points, but the situation can be much worse for some orbital parameters.  Likewise, we examined two other astrophysical noise sources related to the mean-field stellar potential: the mass precession arising from the total stellar monopole moment and the next order of precession arising from the total stellar quadrupole moment.

Even in the limit of zero statistical or measurement error, these sources of astrophysical noise are large and, if left unaddressed, significantly limit future tests of the no-hair theorem.  We showed that the simple functional form of these noise sources (MPD effects, mass precession, and the stellar quadrupole moment) allow us to construct combinations of observable precession angles that exclude astrophysical noise sources, either on an individual basis or in combination with each other.  For example, we have produced one combination of shifts and half-shifts that by construction eliminates all errors associated with both the MPD effect and with mass precession.  We have produced another such combination that by construction eliminates all errors from mass precession and the stellar quadrupole moment.  In principle, although the mathematics is likely laborious, similar algebraic combinations of observables could be made to remove higher-order multipole moments in the stellar potential \citep{KocsisTremaine2015}.  

\blue{We note that, aside from orbital measurements of the S-stars, other tests for the no-hair theorem may exist \citep{Johannsen2016}. If a pulsar is located sufficiently close to Sgr A$^*$, its radio pulses could provide another means to test the no-hair theorem \citep{Wex+1999,Liu+2014}.  An alternative possibility to test the no-hair theorem is by using images of the BH ``shadows'' \citep{Johannsen+2010,Broderick+2014}. For a Schwarzschild BH, the shadow is exactly circular and centered on the BH, and for a rotating BH, the shadow is displaced but remains nearly circular (except for high spin values or large inclination). However, if the no-hair theorem is violated, the shape of the shadow can be significantly different. Indirect constraints on the spin of Sgr A$^*$ may also exist: for example, recent works have noted that several of the innermost S-stars orbit in flattened, disk-like configurations \citep{Ali+2020}.  Such a disk, or disks, could in principle be destroyed (through isotropization of the nodal angles) by differential nodal precession, and indeed, the Lense-Thirring precession time is less than the stellar age for many of these S-stars unless $\chi_\bullet \lesssim 0.1$ \citep{FragioneLoeb2020, FragioneLoeb2022}.  These indirect spin constraints merit further examination, however.  The VRR timescale is even shorter than the Lense-Thirring time at these radii \citep{KocsisTremaine2011}, and VRR is capable of producing dynamically cold, disk-like configurations of heavy stars \citep{KocsisTremaine2015, SzolgyenKocsis2018, PanamarevKocsis2018}, which may explain the apparent survival of these kinematic features even if $\chi_\bullet$ is large.}

\blue{Outside of the Galactic Center, model-fitting to accretion disks can be another way to test the no-hair theorem. This may be done on the shape of relativistically broadened iron lines or on the X-ray thermal continuum spectra \citep{Bambi+2016,Dovciak+2004,Li+2005}. Another approach is using gravitational-wave (GW) measurements \citep{Ryan1997,Glampedakis+2006,Saleem+2021}. Using today's Earth-based LIGO-Virgo-KAGRA detectors, it is challenging to measure an individual object's spins and quadrupole moments in a GW binary. However, in the future, using the space-based {\it LISA} detector, we will be able to detect the GWs from extreme mass ratio inspirals
and calculate the multipole moments of the central BH.  In some cases, combining multiple methods will lead to a more accurate test for the no-hair theorem \citep{Psaltis+2016}.  }

\blue{In this paper, however, we have restricted our attention to tests using S-star orbits.}
The main strength of our simple algebraic approach \red{to testing the no-hair theorem} is that it can in principle be used to eliminate all non-stochastic sources of astrophysical noise. \blue{
While Bayesian methods \citep{Qi2021+} are capable of extracting the parameters of interest from arbitrary combinations of observables (at the cost of finite error from background noise sources), our method allows observers to target future observations to optimally subtract such astrophysical noise and to minimize unnecessary observations.  For example, our calculations suggest that once statistical noise is sufficiently low, observers interested in measuring $Q_{2\bullet}$ do not need to cover the entire orbit, but only to measure precisely the pericenter and the apocenter passages.
} The main weakness of this approach is that we are still discarding some information contained in the relativistic orbits of S-stars. By working with half-shifts, we access more information than is contained in the full-shifts alone, but less than what would exist in a Bayesian approach that compares real observations to a large library of time-dependent orbits.  In the future, it would be interesting to see if these two approaches could be combined in some way to make use of each of their respective strengths.

\begin{acknowledgments}
The authors gratefully acknowledge helpful conversations with Aleksey Generozov and Bence Kocsis\blue{, and we thank the two anonymous referees for their constructive and helpful reports}. YA and NCS acknowledge financial support from  the Israel Science Foundation (Individual Research Grant 2565/19).
\end{acknowledgments}

\bibliography{ms}

\appendix*

\section{Removing the Stellar Quadrupole Moment}
Here we present our derivation of a 7th-order polynomial that can be used to combine the shifts and half-shifts of two stellar orbits in such a way that (i) mass precession and (ii) precesssion from the mean-field stellar quadrupole moment are both precisely removed from calculation of $A_{\rm Q_2}$.

First, we can find the SMBH spin amplitude ($A_{\rm J}$) and its angles in the sky plane ($A,B$):
\begin{subequations}
\begin{align}
\begin{split}
    &\tan(B-\Omega_{1})=\Big[\cos\Omega_{1}(l_{2}e_{1}\cos\Omega_{1}-l_{1}e_{2}\cos\Omega_{2})\\
    &\qquad-\sin\Omega_{1}(l_{1}e_{2}\sin\Omega_{2}-l_{2}e_{1}\sin\Omega_{1})\Big]/\\
    &\qquad\Big[\cos\Omega_{1}(l_{1}e_{2}\sin\Omega_{2}-l_{2}e_{1}\sin\Omega_{1})\\
    &\qquad+\sin\Omega_{1}(l_{2}e_{1}\cos\Omega_{1}-l_{1}e_{2}\cos\Omega_{2})\Big],
\end{split}\\
\begin{split}
    &A_{\rm J}=\frac{3\pi}{e}\frac{1}{\sin i}[k^{2}+\frac{l^{2}}{4}(\tan^{2}(B-\Omega)\cos^{2}i\\
    &\qquad+\frac{\sin^{2}i}{\cos^{2}(B-\Omega)})-lk\tan(B-\Omega)\cos i]^{1/2}
\end{split}
\end{align}
\end{subequations}
where the subscripts are the star index ($A_{\rm J}$ can be calculated using any star) and
\begin{subequations}
\begin{align}
        k_j&\equiv\delta i_{\frac{1}{2},j}-\frac{\delta i_j}{2}-\frac{2e_j}{3\pi}\sin{i_j}\delta\Omega_j,\\
        l_j&\equiv\frac{\sin{i_j}\delta\Omega_j}{2}-\sin{i_j}\delta\Omega_{\frac{1}{2},j}+\frac{2e_j}{3\pi}\delta i_j.
\end{align}
\end{subequations}
The SMBH spin angles in the orbital plane are:
\begin{subequations}
\begin{align}
    &\sin{\alpha_j}\cos{\beta_j}=\frac{3\pi}{2e_j}\frac{l_j}{A_{\rm J}},\\
    &\sin{\alpha_j}\sin{\beta_j}=\frac{3\pi}{e_j}\frac{k_j}{A_{\rm J}},\\
    &\cos{\alpha_j}=\frac{3\pi}{e_j}\frac{k_j}{A_{\rm J}}\cot{i_j}-\frac{3\pi}{2e_j}\frac{l_j}{A_{\rm J}}\tan{(B-\Omega_j)}\sec{i_j}.
\end{align}
\end{subequations}
And finally, to find $A_{\rm Q_2}$ we need to solve a seventh order polynomial:
\begin{equation}
\begin{split}
   & \frac{\Delta\Omega_{1}^{2}\Delta i_{1}^{2}+\Delta\omega_{1}^{2}\Delta i_{1}^{2}+\Delta\Omega_{1}^{2}\Delta\omega_{1}^{2}}{\Delta\omega_{1}\Delta\Omega_{1}\Delta i_{1}}=\\
    &\frac{\Delta\Omega_{2}^{2}\Delta i_{2}^{2}+\Delta\omega_{2}^{2}\Delta i_{2}^{2}+\Delta\Omega_{2}^{2}\Delta\omega_{2}^{2}}{\Delta\omega_{2}\Delta\Omega_{2}\Delta i_{2}}
    \end{split}
\end{equation}
where we denote:
\begin{subequations}
\begin{align}
\begin{split}
    \Delta\Omega_{j}(A_{\rm Q_2})&\equiv A_{\rm J}\sin\alpha_{j}\sin\beta_{j}\\
    &-A_{\rm Q_2}\cos\alpha_{j}\sin\alpha_{j}\sin\beta_{j}-\sin{i}\delta\Omega_{j}
\end{split}\\
\begin{split}
        \Delta i_{j}(A_{\rm Q_2})&\equiv A_{\rm J}\sin\alpha_{j}\cos\beta_{j}\\
        &-A_{\rm Q_2}\cos\alpha_{j}\sin\alpha_{j}\cos\beta_{j}-\delta i_{j}
\end{split}\\
\begin{split}
        \Delta\omega_{\frac{1}{2},j}(A_{\rm Q_2})&\equiv-A_{\rm Q_2}\sin^{2}\alpha_{j}\cos\beta_{j}\sin\beta_{j}\\
        &+\frac{3\pi e_{j}}{2(1+2e_{j}^{2})}(\frac{\delta\omega_{j}}{2}-\delta\omega_{\frac{1}{2},j}).
\end{split}
\end{align}
\end{subequations}

\end{document}